\documentclass[a4paper,11pt]{article}
\usepackage{pos}
\usepackage{multicol,caption}
\usepackage{lipsum}
\usepackage{ulem}

\title{Quarkonium dynamics in the quantum Brownian regime with non-abelian quantum master equations}

\author*[a]{Aoumeur Daddi Hammou}
\author[b]{Stéphane Delorme}
\author[c]{Jean-Paul Blaizot}
\author[a]{Pol-Bernard Gossiaux}
\author[a]{Thierry Gousset}

\affiliation[a]{SUBATECH UMR 6457 , IMT Atlantique, Nantes University, IN2P3/CNRS,\\
4 Rue Alfred Kastler, F-44307 Nantes, France}

\affiliation[b]{Institute of Nuclear Physics, Polish Academy of Science,\\
ul. Radzikowskiego 152, 31-34, Kraków, Poland}
\affiliation[c]{Institut de Physique Th\'{e}orique, Universit\'{e}  Paris Saclay, CEA, CNRS,\\
F-91191 Gif-sur-Yvette, France}
\emailAdd{daddiham@subatech.in2p3.fr}
\emailAdd{stephane.delorme@ifj.edu.pl}
\emailAdd{jean-paul.blaizot@ipht.fr}
\emailAdd{Pol-Bernard.Gossiaux@subatech.in2p3.fr}
\emailAdd{Thierry.Gousset@subatech.in2p3.fr}

\abstract{

Quarkonium production in ultrarelativistic heavy ions collisions is one of the best probes of the QGP formed in these collisions. Resorting to accurate methods to describe the $Q\bar{Q}$
evolution in a QGP is a prerequisite for the precise interpretation of experimental data. Among these methods, the quantum master equations (QME) derived  within the formalism of open quantum systems are particularly  relevant. We present exact numerical solutions in a 1D setting of previously derived quantum master equations (QME)  in  their quantum Brownian regime. Distinctive features of the in-medium bottomonia
evolution with the QME are presented; some phenomenological consequences are addressed by considering evolutions for a fixed as well as  EPOS4 temperature profiles. Next, we investigate the accuracy of the semiclassical approximation (often used to describe charmonium production in URHIC) by benchmarking the corresponding evolutions on the exact solutions derived with the QME for the case of a $c\bar{c}$
pair.}

\FullConference{42nd International Conference on High Energy Physics (ICHEP2024)\\
18-24 July 2024\\
Prague, Czech Republic\\}

\begin{document}
\maketitle

\section{Introduction}
The QME have thus far been derived in the dilute limit (i.e., for a single $Q\bar{Q}$ pair),  rendering them more appropriate for the description of bottomonia dynamics than charmonia, which are produced in large numbers.  The investigation of multiple $c\bar{c}$ pairs  dynamics is currently only possible through the use of semiclassical equations.  The effect of transitioning from an exact quantum description    to an approximate semiclassical description on the precision of the results  remains largely unexplored. 

In these proceedings, following \cite{S.Delorme},   we focus on the quantum Brownian regime of the in-QGP quarkonia dynamics and its associated QME derived in \cite{J-P.Blaizot}. Given that the QME in their dilute limit  are more suitable for the study of bottomonia, we begin by exploring  the capacity of these equations to capture some well-known aspects of the in-medium bottomonia dynamics, particularly  the sequential suppression phenomenon.  In this study, the equations are solved using different initial states and medium configurations.

Subsequently, a comparative study is conducted between the exact quantum results and their semiclassical approximation for the case of a single $c\bar{c}$ in the abelian limit. This study aims to assess the validity of the semiclassical approximation (e.g. used in  \cite{S.Delorme,Denys}) and to provide a stronger foundation for its use in the study of multiple $c\bar{c}$ pairs dynamics.    

\section{Theoretical background}
Starting from the Non-Relativistic QCD (NRQCD) effective theory and working in the quantum Brownian regime of open quantum systems, corresponding to temperatures higher than the quarkonium binding energy, i.e. $T\gg E$, one can derive a non-abelian quantum  master equation of the form \cite{J-P.Blaizot, S.Delorme} 
\begin{equation}
\frac{\text{d}\hat{\rho}_{Q\bar{Q}}\left(t\right)}{\text{dt}}=\hat{\mathcal{L}}\left[\hat{\rho}_{Q\bar{Q}}\left(t\right)\right]=\sum_{i=0}^4\hat{\mathcal{L}}_i\hat{\rho}_{Q\bar{Q}},\label{QME1}
\end{equation}
where $\hat{\rho}_{Q\bar{Q}}$ is the density operator which  describes the quantum state of the $Q\bar{Q}$ pair, and  $\hat{\mathcal{L}}$ is the total Liouville superoperator  that propagates the $Q\bar{Q}$ state in time.   The various  contributions $\hat{\mathcal{L}}_i$ to this superoperator capture different aspects of the dynamics. In particular,  $\hat{\mathcal{L}}_{0,1}$ describe the unitary (or coherent) dynamics with $\hat{\mathcal{L}}_{1}$ encoding the QGP-induced  screening of the real potential. The other superoperators $\hat{\mathcal{L}}_{2\rightarrow4}$, proportional to the so-called imaginary potential,   describe the non-unitary dynamics of the $Q\bar{Q}$ state,  leading eventually to decoherence and thermalization.   It is worth to note that the superoperator $\hat{\mathcal{L}}_{4}$ is subdominant but mandatory for the QME to preserve positivity.  

By performing the projections in color space, we obtain a set of singlet-octet coupled master equations that can be written in the following form 
\begin{equation}
    \frac{d}{dt}\left(\begin{array}{c}
\rho_{\rm s}\\
\rho_{\rm o}
\end{array}\right)=\left(\begin{array}{cc}
\mathcal{L}_{\rm ss} & \mathcal{L}_{\rm so}\\
\mathcal{L}_{\rm os} & \mathcal{L}_{\rm oo}
\end{array}\right)\left(\begin{array}{c}
\rho_{\rm s}\\
\rho_{\rm o}
\end{array}\right),\label{QME}
\end{equation}
where $\mathcal{L}_{\rm so}$ and  $\mathcal{L}_{\rm os}$ describe  the singlet $\leftrightarrows$ octet transitions. Thus, they encode the quarkonium dissociation and recombination.   The equation (\ref{QME}) will be employed  in the next section to study the dynamics of  a single $b\bar{b}$ pair. In this study, we use the one dimensional potentials devised and studied in  \cite{katz}. 

One of the main features of the QME is the possibility of approximating it by a semiclassical equation as shown in \cite{J-P.Blaizot}. 
This approximation  is  based on the assumption that the density matrix   is nearly diagonal. Alternatively, this can be expressed as the system having a short quantum coherence length, beyond which the probability of any superposition state is  suppressed. 
In fact, as discussed in \cite{J-P.Blaizot}, this approximation is best implemented in the abelian or QED version of Eq. (\ref{QME1}), as color to singlet transition imply finite energy exchanges.  We thus restrict ourselves to this case and plan to address the genuine QCD case in a future work. 


The semiclassical  limit of Eq. (\ref{QME1})   was derived in \cite{{J-P.Blaizot}} and  is given by  the following semiclassical Fokker-Planck equation: 
\begin{equation}
    \begin{array}{ccl}
\frac{\partial W\left({r},{p}\right)}{\partial t} & = & \left[-\frac{2{p}{\nabla}_{{r}}}{M}-{\nabla}_{{r}}V\left({r}\right).{\nabla}_{{p}}+\frac{\eta\left({r}\right)}{2}{\nabla}_{{p}}^{2}\right.\\
 &  & \left.+\frac{\gamma\left({r}\right)}{M}{\nabla}_{{p}}{p}\right]W\left({r},{p}\right)\\
\end{array}\label{fokker},
\end{equation}
where $W\left({r},{p}\right)$ is the Wigner distribution that is associated to the density operator  $\hat{\rho}_{Q\bar{Q}}$. It can be shown that the steady state solution of Eq. (\ref{fokker}) is a Gibbs-Boltzmann distribution.  This is also the case for the quantum dynamics as  described by Eq. (\ref{QME1}), but with an effective temperature rather than the real medium temperature. See \cite{S.Delorme} for a discussion of the equilibrium properties associated to the QME.

It should be noted  that in deriving the semiclassical equation (\ref{fokker}) from equation (\ref{QME1}), the operator $\hat{\mathcal{L}}_{4}$ was not included, mainly   due to its subdominance. In principle, this superoperator is only mandatory in the quantum case to preserve the positivity. Therefore, in section  \ref{QM vs. SC}, the semiclassical results will be compared to the quantum ones, both in the presence and absence of this operator, so defining an "acceptability band" for the SC approximation.

\section{Bottomonium quantum dynamics}

Due to their large mass, bottom quarks are not produced in large numbers in the current collisions. As a result, the bottomonia systems in AA collision are naturally governed equation (\ref{QME}), which describes the dynamics of a single $Q\bar{Q}$ pair. 
 
 We initially consider a static QGP at a fixed temperature of 400 MeV and solve Eq. (\ref{QME}) for different initial states in the singlet color channel, namely,  $\Upsilon(1\text{S})$-like, $\Upsilon(2\text{S})$-like, and a mixture of S and P states, i.e. superposition state,  defined  as: 
 \begin{equation}
            \psi(x)=e^-{\frac{x^2}{2\sigma^2}}\left(1+a_{\rm odd}\frac{x}{\sigma}\right),\label{superposition}
\end{equation}
with  $\sigma=0.45$ fm, and  $a_{\rm odd}=3.5$. 

\begin{figure}[htbp]
\begin{center}
\includegraphics[width=16.3cm,keepaspectratio]{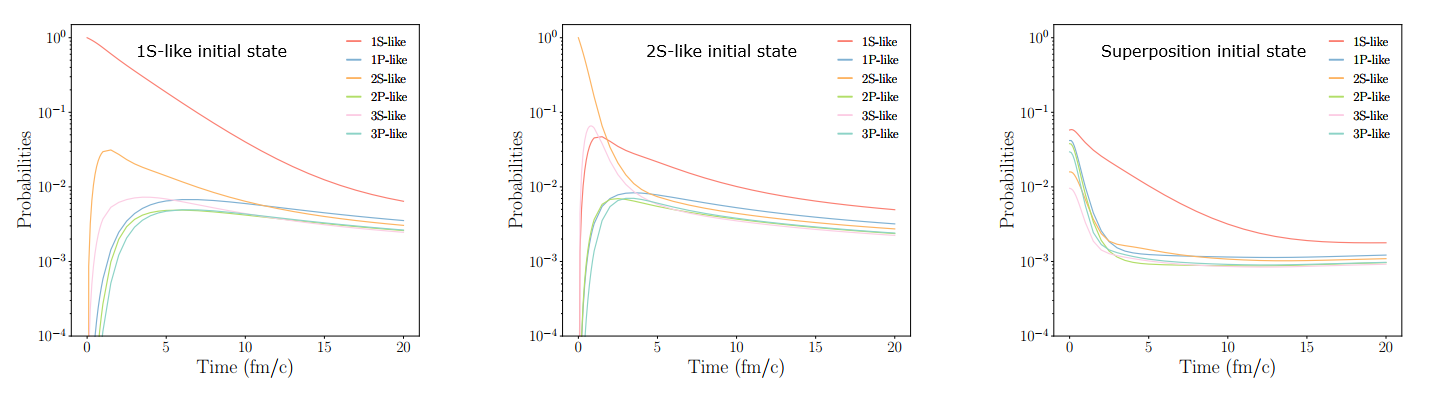}
\end{center}
\caption{Time evolution of bottomonia populations at fixed temperature for different initial states.}
\label{bb-fixed-T}
\end{figure}

Based on the  survival probabilities of the various vacuum eigenstates,  plotted  in Fig. \ref{bb-fixed-T}, we conclude that  the initial state choice has a minimal impact  on the determination of the late-time population values. This is a characteristic of the Markovian dynamics and of the process of thermalization which drives the system towards its equilibrium state. Moreover, the survival probabilities around 10 fm/c in the right panel of Fig. \ref{bb-fixed-T} are indicative of the well-established phenomenon of sequential suppression, whereby the loosely bound states exhibit a greater suppression than the deeply bound states, f.i. Y(1S).  This observation lends support to the reliability of the equation (\ref{QME}) in describing the in-QGP quarkonia dynamics.

\begin{figure}[htbp]
\begin{center}
\includegraphics[width=16.3cm,keepaspectratio]{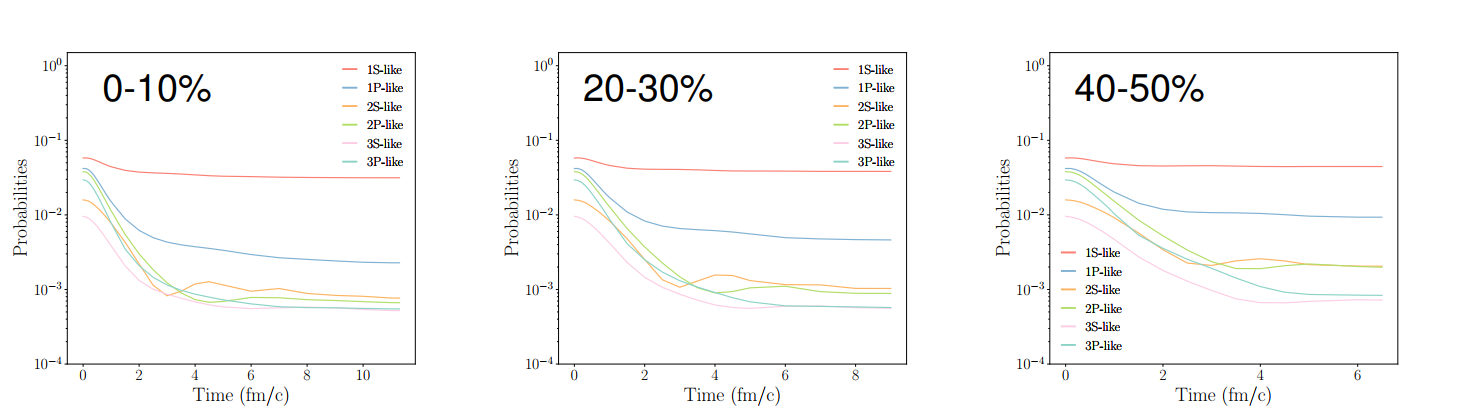}
\end{center}
\caption{Time evolution of bottomonia populations for different centrality classes with EPOS4 averaged temperature profile.}
\label{bb-epos-T}
\end{figure}

We now turn our attention to a dynamical medium which  follows averaged temperature profiles obtained from EPOS4 with the rapidity interval of the CMS experiment, i.e. $|y|<2.4$. In this regard, three distinct centrality classes are considered: 0-10\%, 20-30\%, 40-50\%. The initial state is taken to be the superposition state given by Eq. (\ref{superposition}).  We present in Fig. \ref{bb-epos-T} the survival probabilities  of the various vacuum eigenstates. As expected, we notice a larger suppression in the central collisions compared to peripheral ones. We plan to render this observation more quantitative  by computing the nuclear modification factor using the survival probabilities presented in Fig. \ref{bb-epos-T} as an input.

\section{Quantum vs. semiclassical dynamics of charmonia\label{QM vs. SC}}
We now undertake a comparative analysis between the exact quantum (QM) dynamics, as described by Eq. (\ref{QME1}) in its abelian limit, and the semiclassical (SC) dynamics, as represented by the Langevin equation associated with Eq. (\ref{fokker}). The comparative study focuses on charmonia states, where a semiclassical description is necessary to investigate the dynamics of multiple $c\bar{c}$ pairs. In order to achieve this, the two equations are solved with the same potentials  \cite{katz} and initial state. The latter corresponds to the ground state of our vacuum potential (i.e. 1S-like state) defined as: $\psi\left(x\right)=\left({2}/{\pi\sigma^{2}}\right)^{\frac{1}{4}}e^{-\frac{x^{2}}{\sigma^{2}}}$ with $\sigma=0.54$ fm.

\begin{figure}[htbp]
\centering
\begin{multicols}{3}
\includegraphics[width=5cm,height=2.8cm]{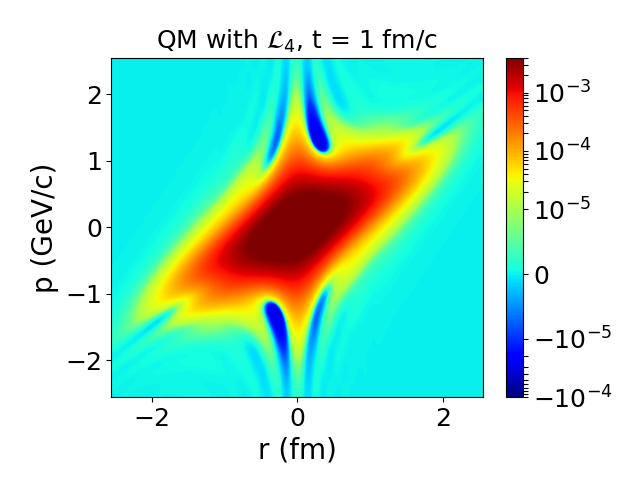}\par

\includegraphics[width=5cm,height=2.8cm]{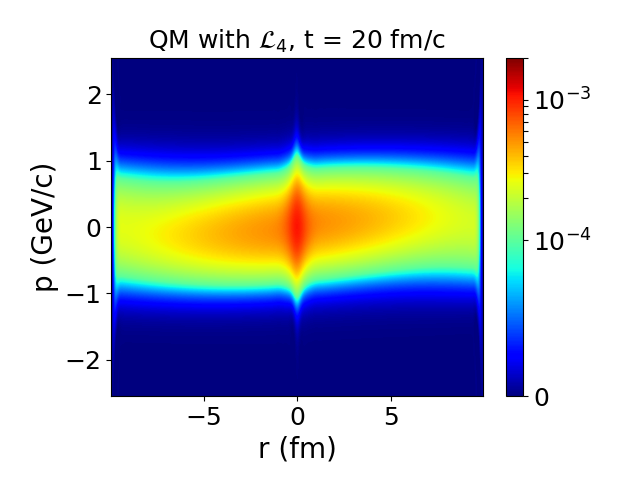}\par

\includegraphics[width=5cm,height=2.8cm]{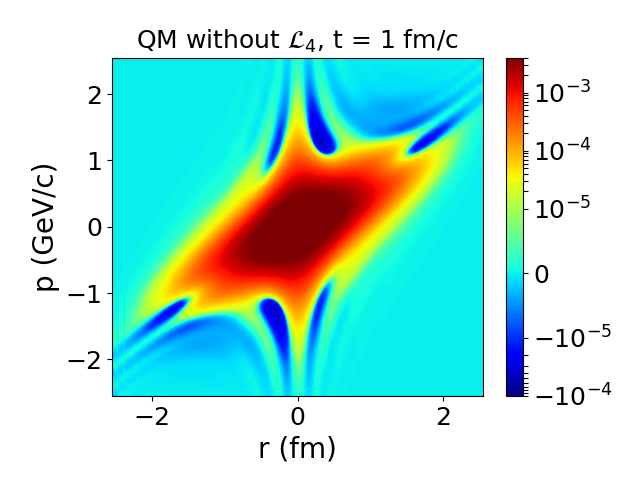}\par

\includegraphics[width=5cm,height=2.8cm]{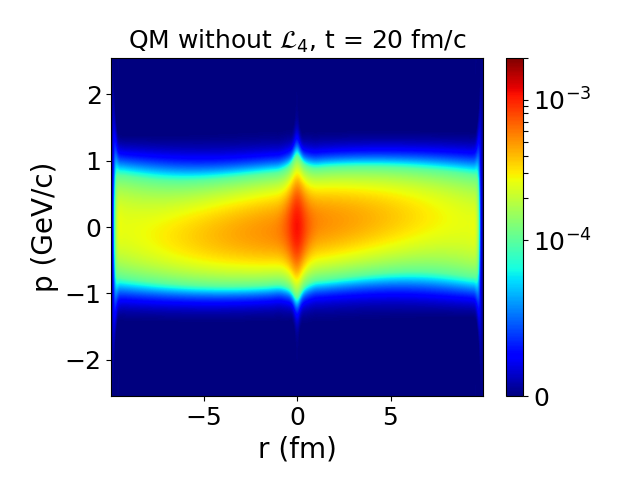}\par
\includegraphics[width=5cm,height=2.8cm]{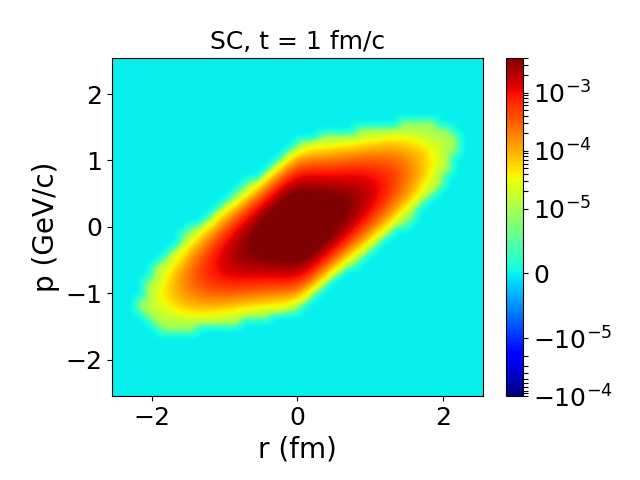}\par

\includegraphics[width=5cm,height=2.8cm]{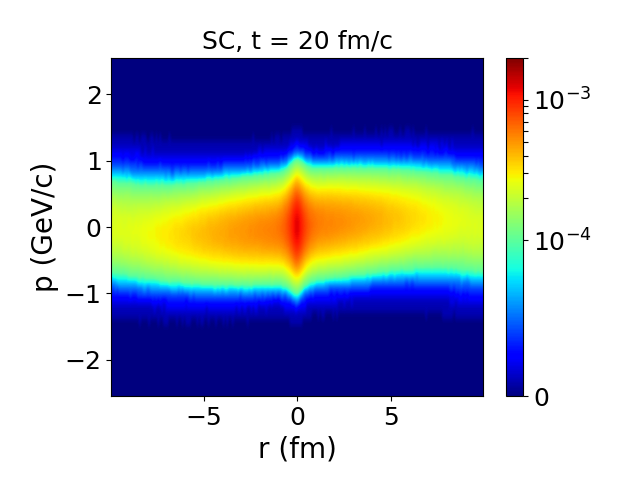}\par
\end{multicols}
\caption{Quantum (QM) vs. semiclassical (SC) time evolution of the Wigner distribution resulting from the evolution of a 1S-like state in a QGP at $T=0.3$~GeV.}\label{fig:snapshots-wigner}
\end{figure}

In Fig. \ref{fig:snapshots-wigner},   we compare  the quantum and semiclassical Wigner distributions  at different times.  It can be observed that at early times, discrepancies emerge due to the interference effects inherent to the quantum description but absent in the semiclassical case. However, once quantum decoherence suppresses these interference effects, a better alignment between the quantum and semiclassical descriptions is observed at intermediate and late times.

In order to render this comparison more precise, we compute the trace distance between the quantum and semiclassical Wigner distributions, defined as:
\begin{equation}
    \text{d(t)}=\sqrt{2\pi\hbar\int\text{drdp}\left(W_{QM}(t)-W_{SC}(t)\right)^{2}},\label{eq:trace-distance-def}
\end{equation}
and bounded by 2. Thus, half of the distance can be thought of as a percentage of discrepancy between the two descriptions, i.e.   ${\text{d}(t)}/{2}\in [0,100]\%$. Given that the initial state is the same for both equations, the distance vanishes at the outset. 

\begin{figure}[htbp]
\centering
\includegraphics[width=0.58\textwidth]{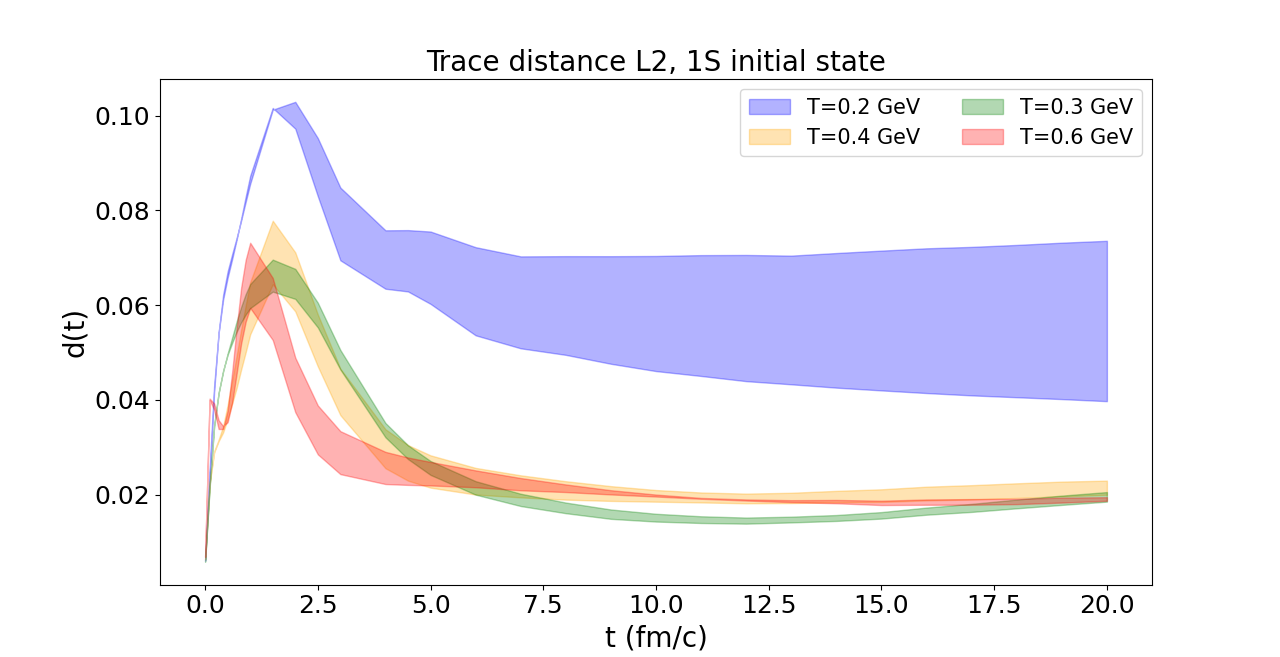}
\caption{Time evolution of the distance  given by ~Eq.~(\ref{eq:trace-distance-def}) for various QGP temperatures. The band encodes   the uncertainty related to the presence or absence of $\hat{\mathcal{L}}_{4}$ in the quantum dynamics. }
\label{fig:trace-dist}
\end{figure}

Figure. \ref{fig:trace-dist} depicts the temporal evolution of the distance for a range of QGP temperatures. It is observed that the maximum distance value is  $\text{d(t)}\simeq0.1$, which  corresponds to T=0.2 GeV. This indicates that there is at least  $\sim$95\% of agreement between the two descriptions. This percentage is even larger for higher temperatures and/or at late times. In light of these findings, it can be concluded that the semiclassical description provides an accurate approximation of the exact quantum description, while offering a more computationally efficient alternative. 

\begin{figure}[htbp]
    \centering
    \includegraphics[width=0.49\textwidth]{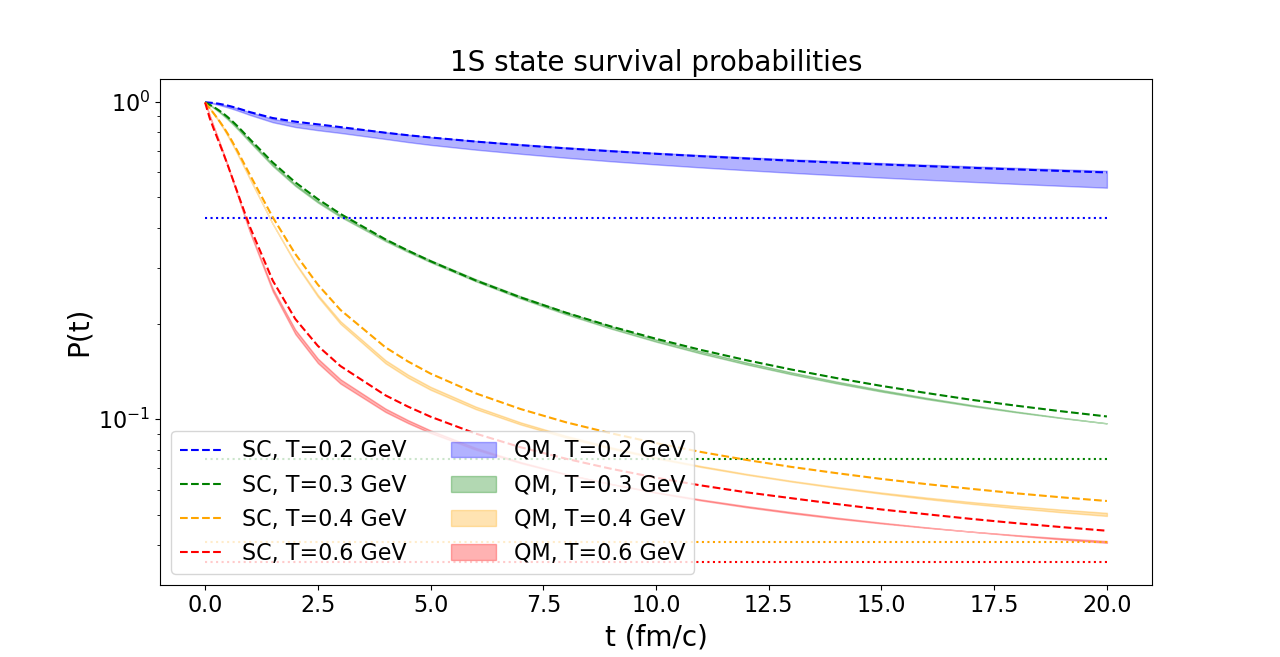}
     \includegraphics[width=0.49\textwidth]{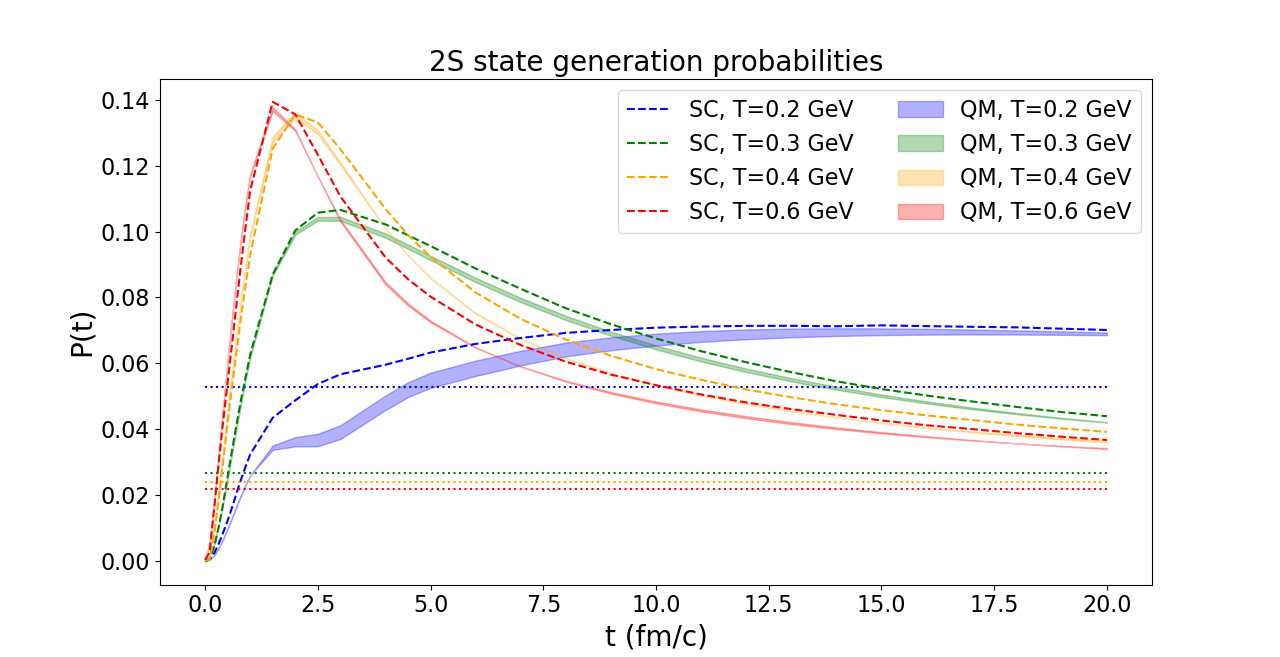}
    \caption{QM vs. SC survival probability of the 1S-like state (left) and generation probability of the 2S-like state (right). The dotted lines are the equilibrium values expected from the Gibbs-Boltzmann distribution.}
    \label{fig:1S-state-problt}
\end{figure}

To further examine the reliability of the semiclassical approximation, we calculate the probabilities of survival for the 1S-like state and the generation of the 2S-like state. The results are presented in Fig. \ref{fig:1S-state-problt}. The semiclassical results are in close agreement with the exact quantum results, with the exception of a slight discrepancy observed at mid and late times which is due to the different equilibrium states in the two descriptions.

\section{Conclusion}

We have explored the in-QGP quarkonia dynamics through the lens of  the quantum master equations approach. In particular, we investigated the capacity of this approach to describe the various aspects related to the QGP-induced suppression of quarkonia population. Moreover, a quantitative assessment of the semiclassical approximation was conducted, demonstrating its capability to accurately reproduce the exact quantum results.

\end{document}